\documentclass[pra,aps,showpacs]{revtex4}

\usepackage{graphics}

\begin{document}
\draft
\title{{\large {\bf Two-photon coherent control of atomic collisions by light
with entangled polarization}}}
\author{M.D.Havey}
\address{Physics Department Old Dominion University\\
Norfolk, Virginia 23529}
\author{D.V.Kupriyanov and I.M.Sokolov}
\address{Department of Theoretical Physics, State Technical University, 195251,
St.-Petersburg, Russia}
%\date{\today}

\begin{abstract}
We describe a new method of coherent optical control of internal dynamics of
atomic collisions by means of two correlated light beams having entangled
polarizations. We show that if excitation of a colliding pair of atoms is by
two photons having entangled polarizations, it is possible to redirect the
output fragments of the collision into certain channels with a selected type
of internal transition symmetry. The transition symmetry is defined in the
body-fixed coordinate frame which has random and originally unknown
orientation in space.
\end{abstract}


%% 34.50.Rk, 34.80.Qb - Laser modified scattering and reactions
%% 42.50.Ct - Quantum description of interaction of light and
%%            matter, related experiments
\maketitle

\noindent Significant experimental developments over the past decade have
led to remarkable understanding of details of atomic collision dynamics.
Studies of a broad spectrum of processes, ranging from cold and ultracold
collisions \cite{Rev1,Rev2,Rev3}, photoassociation \cite{Bla,Jones},
photodissociation \cite{Zaf,ABS}, photochemical reactions \cite{Zwl,Femto},
optical and fractional collisions \cite{OHSL,HBOT}, and collisional
redistribution of light \cite{RevB} have revealed novel and often surprising
effects depending on variables associated with the collision alone and with
properties of the light used to initiate or probe the dynamics. In many
cases, dynamical correlation of internal variables of the colliding
particles have played a critical role in the outcome; for light-induced
processes, dependencies on the cross sections due to classical
characteristics of the light, viz. polarization, frequency and intensity
have been determined. \ A novel method to obtain coherent control of the
correlations in photodissociation, by using elliptically polarized light,
demonstrated selectivity in the branching ratios for the process\cite{ABS}.
\ Further, general principles for coherent control of collision and reactive
processes using a single light source have recently been developed in a
fundamental paper on control of bimolecular scattering processes \cite
{Shapiro}. \

Due to recent advances in studies of photochemical processes on a
femtosecond time scale \cite{Zwl,Femto} and, as was recently shown in atomic
collision experiments \cite{OHSL,HBOT}, it becomes possible to optically
probe a colliding system directly in the interaction domain and to select in
this manner a small segment of a collision trajectory. Such a process,
termed a fractional optical collision, is an example of a
continuum-continuum two-photon spectroscopy where the first and the second
photons are used for initiating and  interrupting the collisional motion in
an intermediate molecular state. In studies up to now, photoexcitation was
driven by two independent light sources used mainly for selecting and
probing the location of the Condon points of the fractional collision as
described by quasistatic conditions of photoexcitation. The details of the
internal collisional dynamics, as well as the selective information about
different channels involved in the process were difficult to extract from
the data obtained from the spectroscopic analysis, in spite of the fact that
both intensity and polarization spectra were determined.

In the present letter, we describe how the outcome of a fractional collision
may be significantly and selectively controlled by utilization of
quantum-correlated light beams. Our approach is based on the requirement
that the photon correlations must interfere with correlations arising from
internal collisional dynamics. In our description of the light statistics we
employ a fully quantal approach, which permits us to discuss the difference
in predictions for classical and quantum electrodynamics. One aim of the
present paper is to point out that to understand precise optical control of
elementary processes like atomic collisions, it is important to follow the
transformation of quantum correlations (existing on a wave function level)
from an electromagnetic subsystem into a diatomic (or multi-atomic)
subsystem. As a practical example of non-classical light, we consider the
radiation from an optical parametric oscillator (OPO) operating in a
subthreshold regime, and so having entangled polarizations of the output
modes. Different schemes of practical realization of entangled states with
the aid of optical parametric oscillators have been discussed in the
literature \cite{ReDr,OPKP,YuSt,ZZHE,SSRKA,KMW}.

In a perturbation theory approach any two-photon process can be described in
terms of the light correlation function of second order, i.e. in terms of
time-($T$) and anti-time-($\tilde{T}$) ordered products of the Heisenberg
operators of positive and negative frequency components of electric field
amplitudes $E_{\nu }^{(\pm )}({\bf r}t)$ considered as functions of space ($%
{\bf r}$) and time ($t$) coordinates \cite{Glb}. For a two-mode OPO output
with entangled orthogonally polarized components, such a correlation
function can be expanded in the following sum
\begin{eqnarray}
D_{\nu_1\nu_2;\nu^\prime_1\nu^\prime_2}(\tau)&=&
\langle{\tilde T}(\!E^{(-)}_{\nu'_1}({\bf r}t)\,
E^{(-)}_{\nu'_2}({\bf r}t+\tau))\,
T(\!E^{(+)}_{\nu_1}({\bf r}t)\,
E^{(+)}_{\nu_2}({\bf r}t+\tau))\rangle=
\nonumber\\
&&\sum\limits_{1,2,1',2'}
({\bf e}_1)_{\nu_1}\,({\bf e}_2)_{\nu_2}\,
({\bf e}_1'^*)_{\nu_1'}\,({\bf e}_2'^*)_{\nu_2'}\:
D_{121'2'}(\tau)
\label{1}\end{eqnarray}

where each of the polarization vectors ${\bf e}_{i},\,{\bf e}_{i}^{\prime }$
(with $i=1,2$) is one of the basic orthogonal polarizations of the OPO. The
sum over $1,2,1^{\prime },2^{\prime }$ is restricted by the rule ${\bf e}%
_{1}\neq {\bf e}_{2}$ and ${\bf e}_{1}^{\prime }\neq {\bf e}_{2}^{\prime }$,
so there are four terms in the expansion (\ref{1}). We assume here steady
state and homogeneous conditions of photoexcitation and consider the
correlation function only as a function of the time delay between
appearances of the first and second photons.

Strictly speaking the above expansion of the full correlation function
relates to the limit of weak sub-threshold OPO source, generating the photon
pairs, see Eq.(\ref{8}). This is the most interesting and important case for
our discussion. But in a more general situation, to introduce the expansion (%
\ref{1}), we need to cancel out the non-correlated contribution when both
the photons appear in the same polarization mode. However even in a general
situation, for methodical clarity, it is useful to discuss the correlation
function in form (\ref{1}) since it lets us compare the difference between
quantum and classical types of polarization entanglement.

A schematic diagram illustrating the process of two-photon excitation of
colliding atoms is shown in Figure 1. There the vertical lines represent
optical transitions, while the paths along the interatomic potentials
indicate the kinetic motion of the colliding atoms. Based on the
Franck-Condon approximation and on the assumption of adiabatic evolution of
the diatomic system in the intermediate states, the total cross-section (or
transition probability) of the fractional collision can be expressed as
follows
\begin{eqnarray}
\sigma_0&=&
\sum\limits_{1,2,1',2'}
\sum\limits_{X\Xi}\, (-)^{X+\Xi}\,
\Phi_{X\Xi}({\bf e}_1,{\bf e}_1'^*)\,
\Phi_{X-\Xi}({\bf e}_2,{\bf e}_2'^*)\times
\nonumber \\
&&\phantom{\sum\limits_{1,2,1',2'}
\sum\limits_{X\Xi}}
Q_{121'2'}^{(X)}
\label{2}\end{eqnarray}
where the tensor functions
\begin{eqnarray}
\Phi_{X\Xi}({\bf e},{\bf e}'^*)&=& -\sum\limits_{\nu,\nu'}\,
{\rm C}_{1\nu'\,1\nu}^{X\Xi}\, ({\bf e}'^*)_{\nu'}\,e_\nu=
\nonumber \\
&&\sum\limits_{\nu,\nu'}\,(-)^{1+\nu'}\,
{\rm C}_{1\nu'\,1\nu}^{X\Xi}\,e_{-\nu'}'^*\,e_\nu\, ,
\label{3}\end{eqnarray}
considered as a function of ${\bf e},{\bf e}^{\prime }=$ either ${\bf e}_{1},%
{\bf e}_{1}^{\prime }$ or ${\bf e}_{2},{\bf e}_{2}^{\prime }$, are the
irreducible polarization components of the OPO light. Here by ${\rm C}%
_{\ldots \,\ldots }^{\ldots }$ we denote the Clebsch-Gordan coefficients in
the notation of Ref.\cite{VMK}. Each partial contribution of the $X$-rank
components in the irreducible product in Eq.(\ref{2}) is weighted with the
factor $Q_{121^{\prime }2^{\prime }}^{(X)}$ given by
\begin{eqnarray}
Q_{121'2'}^{(X)}&\sim&
\frac{1}{(2j_0+1)(2X+1)}
\sum\limits_{\bar{\Xi}}\sum\limits_{\bar{\nu}_1\bar{\nu}_1'}
\sum\limits_{\bar{\nu}_2\bar{\nu}_2'}\,
(-)^{\bar{\nu}_2'+\bar{\nu}_1}\,
{\rm C}_{1\bar{\nu}_2\,1-\bar{\nu}_2'}^{X\bar{\Xi}}\,
{\rm C}_{1-\bar{\nu}_1\,1\bar{\nu}_1'}^{X\bar{\Xi}}\times
\nonumber \\
&&w^{(1)}\,w^{(2)}
\left[{\rm d}_{\bar{\Xi}\bar{\Xi}}^{X}(\xi_{+-})\:
{\cal D}_{121'2'}(\tau_{+-})\,+\,
{\rm d}_{\bar{\Xi}\bar{\Xi}}^{X}(\xi_{-})\:
{\cal D}_{121'2'}(\tau_{-})\:\theta(R_1-R_2)\,+\right.
\nonumber \\
&&\left.
\phantom{w^{(1)}\,w^{(2)}\;[
{\rm d}_{\bar{\Xi}\bar{\Xi}}^{X}(\xi_{+-})\:
{\cal D}_{121'2'}(\tau_{+-})\;+\;}
{\rm d}_{\bar{\Xi}\bar{\Xi}}^{X}(\xi_{+})\:
{\cal D}_{121'2'}(\tau_{+})\:\theta(R_2-R_1)\right]
\label{4}\end{eqnarray}
where
\begin{equation}
{\cal D}_{121'2'}(\tau)=
(Tr\,D(\infty))^{-1}\,D_{121'2'}(\tau)
\label{5}\end{equation}
is the dimensionless correlation function normalized according to its
classical limit. With reference to Fig. 2., the following notation is used
in Eq.(\ref{4}). First, $j_{0}$ is the angular momentum of the lower state.
In the arguments of the Wigner $d-$ function $\xi _{+-}$ and $\xi _{\pm }$
are the average deflection angles defined for different segments of the
collisional trajectory crossing the Condon points $R_{1}$ and $R_{2}$, as
defined in Fig. 2. $\tau _{+-}$ and $\tau _{\pm }$ are the average durations
of the fractional collision defined for these segments of the trajectory,
while $w^{(1)}$ and $w^{(2)}$ are the Franck-Condon transition probabilities
for optical excitation near the points $R_{1}$ and $R_{2}$ respectively. The
step $\theta $-functions in Eq.(\ref{4}) indicate that such transitions are
acceptable either on incoming or outgoing parts of the motion.

The partial cross section $Q_{121^{\prime }2^{\prime }}^{(X)}$ is the most
important characteristic of the fractional collision process. As follows
from Eqs.(\ref{1}), (\ref{2}), this quantity describes both the total
probability for and the polarization dependence of the process. The
expression (\ref{4}) can be consistently derived based in the general theory
of fractional optical collisions \cite{KSST}. Here we present only a
qualitative description. First, we point out that all the tensor components
in the sum of expression (\ref{4}) relate to the internal molecular
(body-fixed) frame; the corresponding tensor indices are indicated by over
bars. The Clebsch-Gordan coefficients in this expression can be treated as
irreducible components of the light as defined in the molecular frame.
Second, the sum over tensor indices is non-invariant here and is expanded
only over those transitions which are permitted by the Franck-Condon
principle. The Wigner $d$-functions describe the rotational and adiabatic
transformation of the light irreducible components due to adiabatic
evolution of the electronic subsystem during the internal part of the
collision for atomic motion from point $R_{1}$ to point $R_{2}$.

The correlation function of the light appears in Eq. \ref{4} as a function
of fractional collision time intervals, which reveals how the photon
correlations interfere with the collisional dynamics. Consider the situation
when the photons, emitted by a subthreshold OPO, can appear in two
orthogonal polarizations along $x$ and $y$ axes: ${\bf e}_{x}$ and ${\bf e}%
_{y}$. Then, assuming quasi-Gaussian statistics in the averaged product (\ref
{1}), we obtain the following set of dimensionless correlation functions:
\begin{eqnarray}
{\cal D}_{xyxy}(\tau)&=&{\cal D}_{yxyx}(\tau)=
\frac{1}{2}+\frac{1}{2}\coth^2\!\kappa\:g(\tau)
\nonumber \\
{\cal D}_{xyyx}(\tau)&=&{\cal D}_{yxxy}^*(\tau)=
\frac{1}{2}{\rm e}^{-i\varphi}\,\coth^2\!\kappa\:g(\tau)
\label{6}\end{eqnarray}
where $\varphi$ is a phase mismatch between the anomalous correlation
functions in the product
$$
\langle \tilde{T}\,E_{1y}^{(-)}({\bf r}t)\,
E_{2x}^{(-)}({\bf r}t+\tau)\rangle
\langle T\,E_{2y}^{(+)}({\bf r}t+\tau)\,
E_{1x}^{(+)}({\bf r}t)\rangle
$$
Here in the indices we display both the mode number and the polarization. In
our simple model of a subthreshold OPO, $\sinh ^{2}\!\kappa $ is of order of
the number of photons emitted by the crystal in the coherence volume of
parametric radiation. The opposite limits, when the efficiency of the
process $\kappa \rightarrow 0$, or when $\kappa >1$, describe the weak and
strong output respectively. Time correlation of twin-photons is described by
the function $g(\tau )$ which can be controlled with an optical delay line.

From the point of view of classical electrodynamics only the first term
(i.e. $\frac{1}{2}$) in the expression for the diagonal components of the
correlation functions (\ref{6}) is acceptable. The non-classical behavior of
the correlation functions (\ref{6}) can be clearly seen in the weak
radiation limit: if $\kappa \rightarrow 0$ then $\coth ^{2}\!\kappa
\rightarrow \infty $. Actually, such a singularity means that for weak OPO
light the correlation function (\ref{1}) has linear (not quadratic)
dependence on mode intensity. In this case the dimensionless correlation
function, normalized in accordance to (\ref{5}), should approach infinity
for short time delay. The linear dependence of two-photon absorption on
light intensity for the radiation created in the down-conversion process was
recently observed in experiment \cite{GPEKP}. For our discussion it is more
important to point out the non-classical behavior of the polarization for
OPO light. Indeed, in the limit of weak output the two-mode light
illuminating the colliding atoms describes the photon pairs with the
following cooperative wave function:
\begin{equation}
|\Psi\rangle_{12}=\frac{1}{\sqrt{2}}
\left[|{\bf e}_x\rangle_1\,|{\bf e}_y\rangle_2\ +
\ {\rm e}^{i\varphi}\,
|{\bf e}_y\rangle_1\,|{\bf e}_x\rangle_2\right]
\label{8}\end{equation}
For such a polarization-entangled wave function in the case of $\varphi =\pi$
there is no particular polarization for each photon, but there is a strong
mutual orthogonal polarization between them.

As a particular application, consider the collisional system, often
discussed in optical collision theory, for which the optical transitions are
initiated between singlet states of one atom. The second atom (an inert-gas
atom) conserves its electronic configuration during the collision. If the
optically active atom is originally in the ground ${}^1\!S$-state there are
the following dipole-allowed two-photon transitions available: ${}^1\!S \to{}%
^1\!P \to {}^1\!S,\, {}^1\!D$. For simplicity, let us ignore rotational
effects and assume that the main contribution to the transition probability
comes from the recoil collision with small impact parameters. Then we can
substitute expressions (\ref{6}) into Eqs.(\ref{4}) and (\ref{2}) and look
at the partial contribution for each possible pair of Franck-Condon
transitions. In the Franck-Condon approximation we can select all the
acceptable transitions in terms of their molecular symmetry. Also in the
recoil limit we can ignore the difference between incoming-outgoing and
either incoming or outgoing parts of the classical trajectory, since all of
them have similar polarization dependence.

With such assumptions the partial contribution to the cross-section for the
excitation via ${}^{1}\!\Sigma \rightarrow {}^{1}\!\Sigma \rightarrow
{}^{1}\!\Sigma $ or ${}^{1}\!\Sigma \rightarrow {}^{1}\!\Pi \rightarrow
{}^{1}\!\Sigma $ is given by
\begin{equation}
\sigma_0\propto
\frac{1}{15}\,w^{(1)}\,w^{(2)}
\left[1+(1+\cos\varphi)\:\coth^2\!\kappa\:g(\tau)\right]
\label{9}\end{equation}
For the excitation via ${}^{1}\!\Sigma \rightarrow {}^{1}\!\Sigma
\rightarrow {}^{1}\!\Pi $ or ${}^{1}\!\Sigma \rightarrow {}^{1}\!\Pi
\rightarrow {}^{1}\!\Pi $ we obtain
\begin{equation}
\sigma_0\propto
\frac{4}{15}\,w^{(1)}\,w^{(2)}
\left[1+(1-\frac{1}{4}\cos\varphi)\:
\coth^2\!\kappa\:g(\tau)\right]
\label{10}\end{equation}
and the contribution for the excitation via ${}^{1}\!\Sigma \rightarrow
{}^{1}\!\Pi \rightarrow {}^{1}\!\Delta $ channel is given by
\begin{equation}
\sigma_0\propto
\frac{1}{5}\,w^{(1)}\,w^{(2)}
\left[1+(1+\cos\varphi)\:\coth^2\!\kappa\:g(\tau)\right]
\label{11}\end{equation}
The dependence of these expressions on phase $\varphi $ reveals how the
quantum correlations, existing between the OPO mode polarizations, can
interfere with the internal dynamics of the fractional collision process.
This is most easily seen in the limit of weak radiation, when the second
terms in the brackets of Eqs.(\ref{9})-(\ref{11}) give dominant
contributions. If the phase $\varphi =\pi $, the transition probability for $%
{}^{1}\!\Sigma \rightarrow {}^{1}\!\Sigma \rightarrow {}^{1}\!\Sigma $, $%
{}^{1}\!\Sigma \rightarrow {}^{1}\!\Pi \rightarrow {}^{1}\!\Sigma $ and $%
{}^{1}\!\Sigma \rightarrow {}^{1}\!\Pi \rightarrow {}^{1}\!\Delta $
excitation channels drops sharply. Such behavior can be understood, based on
the wave function (\ref{8}), where the entangled states have unknown
polarization for each photon, but there is a strong mutual correlation
between their polarizations. If $\varphi =\pi $ and the first photon
possesses unknown polarization along an arbitrary direction in the space
then the second photon has polarization orthogonal to this direction. In
such a case, the absorption of the first photon during the collision fixes
the polarization direction for the second photon, i.e. reduces the
uncertainty of its quantum state. Thus, when $\varphi =\pi $, the second
photon has orthogonal polarization to the direction of the transition dipole
moment. Then it cannot be absorbed in the above examples of the
Franck-Condon transitions and such excitation channels become closed. But at
the same time, the excitations via ${}^{1}\!\Sigma \rightarrow
{}^{1}\!\Sigma \rightarrow {}^{1}\!\Pi $ and ${}^{1}\!\Sigma \rightarrow
{}^{1}\!\Pi \rightarrow {}^{1}\!\Pi $ channels are open and we obtain in
contrast an increase in the transition probability. Let us emphasize here
that, from the point of view of classical electrodynamics, this effect is
forbidden, since it would be impossible to prepare light polarized along (or
orthogonal to) an arbitrary and originally unknown direction in\ space.

In summary, we have shown that unique manipulations of colliding atoms by
light with entangled polarization can result in coherent control of
elementary processes such as atomic collisions or chemical reactions. Even
after full averaging the polarization entanglement makes it possible to
close one channel of a photochemical reaction and to open another if they
have different optical transition symmetries.

Support of this research by the U.S. Civilian Research and Development
Foundation, the National Science Foundation and the Russian Foundation for
Basic Research is gratefully acknowledged. D.K. would like to acknowledge
the financial support from Delzell Foundation, Inc.

\begin{figure}[tbp]
\includegraphics{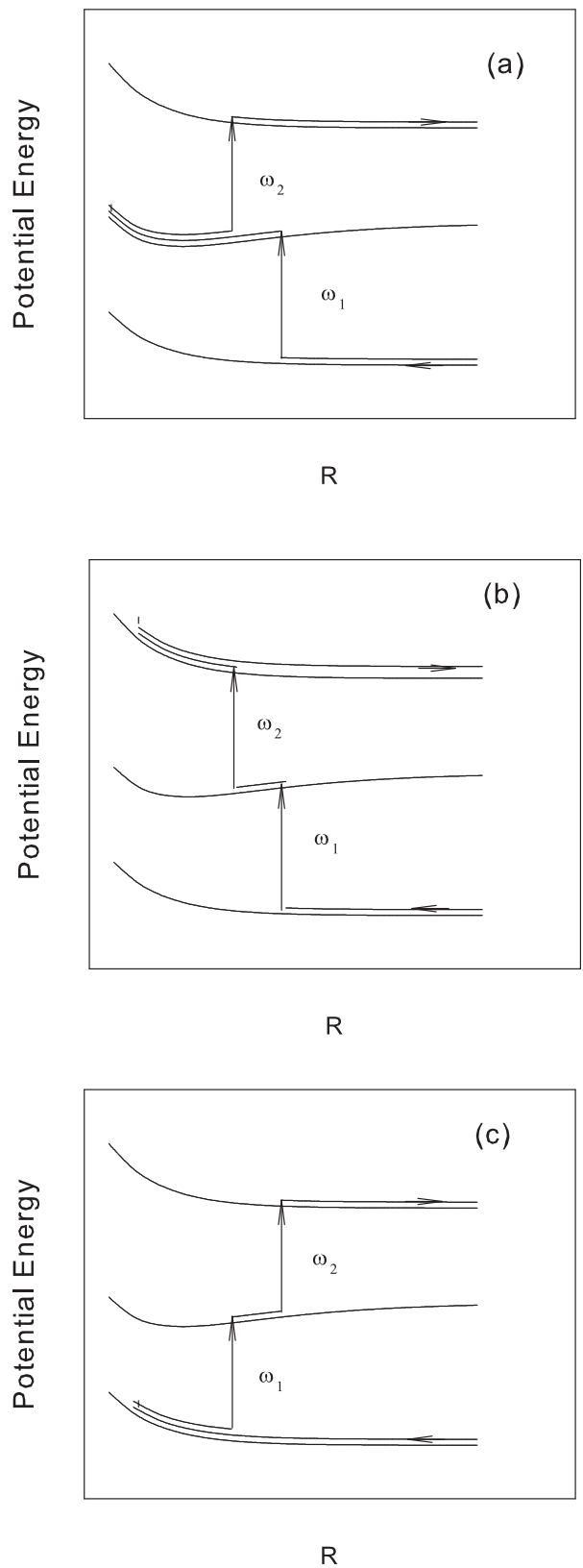}
\caption{Two-photon excitation of colliding atoms for locations of the
Condon points on (a) incoming-outgoing, (b) incoming and (c) outgoing parts
of a classical trajectory.}
\end{figure}

\begin{figure}[tbp]
\includegraphics{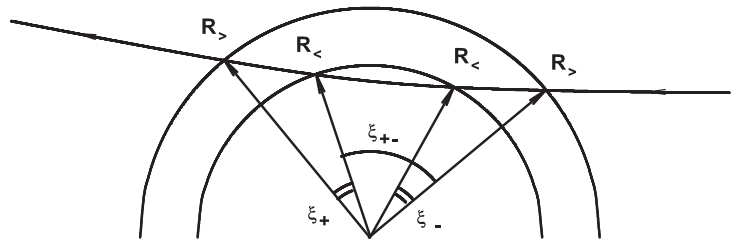}
\caption{Definition of deflection angles for different segments of a
collisional trajectory. Here $R_>$ or $R_<$ is either $R_1$ or $R_2$
depending on the situations shown in Fig.1.}
\end{figure}

\end{document}